# PRELIMINARY: Proton Spectrum at the Jupiter Laser Facility of LLNL

Carol Scarlett[a], Hui Chen[b], Jerry Peterson[c]
[a]Florida Agriculture and Mechanical University
[b]Lawrence Livermore National Laboratory
[c]University of Colorado, Boulder

**ABSTRACT:**
This paper looks at tungsten samples irradiated by beams of protons, gammas, electrons and positrons at the Jupiter Laser Facility of Lawrence Livermore National Laboratory (LLNL). The resulting unstable nuclei created are identified using their gamma spectra. These spectra were taken, usually within an hour of irradiation, for periods up to 48 hrs. In several cases there are two isotopes, one of Rhenium and the other of Tantalum, that emit the same gamma lines. These pairs often involve a long-lived and a short-lived candidate. Spectra were taken 80 days after initial exposure and the long-lived candidates are ruled out.

## I. Introduction:

Researchers at LLNL's Jupiter Laser Facility have demonstrated the largest production of laser induced antimatter particles, in the form of positrons[1]. We are proposing a search for the weak interaction cross section of positrons (e+) and neutrons (n). This process, a form of beta decay, can currently be accessed only through its inverse form of neutrino capture by a proton[2]. Given the quark substructure of the neutron and indications of matter-antimatter interaction anomalies[3], a measure of this cross section may provide evidence of T-Reversal Violations. The work described here represents the first of a series of steps necessary to realize an e+ on n experiment. Here we shall focus on characterizing the spectrum for all particles produced following the illumination of a primary target. An analysis of the raw spectrum, needed to estimate the numbers of each isotopes produced is presented. The next step is to extract the spectra for all types of particles in the secondary beam. Ultimately, the spectrum will be used to calculate the backgrounds that the proposed beta decay experiment will encounter.

Section II describes the methodology of using radioactivity techniques to determine the spectrum of an unknown particle beam (protons, gammas, electrons, etc.). Section III covers the extraction of raw counts from a spectrum taken at LLNL's counting house for the tungsten sample. This section also contains a snapshot of the spectrum taken approximately 80 days later at Florida A&M University (FAMU). Section IV describes what can be learned from comparison of these spectra and concludes which isotopes are consistent with both sets of data.

## II. Determining The Proton Spectrum:

The use and production of particle beams has lead to sophisticated techniques for measuring energy spectra. One such technique involves the exposure of heavy metals to beams with energies exceeding a few million electron volts (MeV)[4-5]. These nuclei often have numerous isotopes that can be created in processes where one or more nucleons are ejected from the target.

Equation (1) shows some sample radioactive nuclei created through interactions between protons and a tungsten target:

$$^{184}W + p \rightarrow {}^{184}Re + n$$
$$^{184}W + p \rightarrow {}^{183}Ta + 2p \quad (1)$$
$$^{186}W + p \rightarrow {}^{184}Ta + n2p$$

Each of these interactions has a unique energy threshold, above which the cross section rises to a plateau. These thresholds can provide a shortcut to placing an upper limit on the energy of protons by simply looking for the absence of certain isotopes. In this note we will use these thresholds to extract the proton energy spectrum.

Table 1 lists the important isotopes and their abundance in naturally occurring Tungsten (W), while Table 2 lists some of the radioactive isotopes that can be produced through interactions with protons.

Table 1: Tungsten Isotopes

| Isotope | Abundance (%) |
|---|---|
| $^{182}W$ | 26.50 |
| $^{183}W$ | 14.31 |
| $^{184}W$ | 30.64 |
| $^{186}W$ | 28.43 |

Table 2: Proton interactions with Tungsten

| Nuclide | Half-life | Contributing Reactions | Threshold Energy (MeV) |
|---|---|---|---|
| $^{181}Re$ | 19.9 h | $^{182}W(p,2n)$ | 10.64 |
| | | $^{183}W(p,3n)$ | 16.87 |
| | | $^{184}W(p,4n)$ | 24.32 |
| | | $^{186}W(p,6n)$ | 37.34 |
| $^{182m}Re$ | 12.7 h | $^{182}W(p,n)$ | 3.60 |
| | | $^{183}W(p,2n)$ | 9.83 |
| | | $^{184}W(p,3n)$ | 17.28 |
| | | $^{186}W(p,5n)$ | 30.30 |
| $^{182g}Re$ | 64.0 h | $^{182}W(p,n)$ | 3.60 |
| | | $^{183}W(p,2n)$ | 9.83 |
| | | $^{184}W(p,3n)$ | 17.28 |
| | | $^{186}W(p,5n)$ | 30.30 |
| $^{183}Re$ | 70.0 d | $^{183}W(p,n)$ | 1.35 |
| | | $^{184}W(p,2n)$ | 8.80 |
| | | $^{186}W(p,4n)$ | 21.82 |
| $^{184m}Re$ | 169 d | $^{184}W(p,n)$ | 2.28 |
| | | $^{186}W(p,3n)$ | 15.29 |
| $^{184g}Re$ | 38.0 d | $^{184}W(p,n)$ | 2.28 |
| | | $^{186}W(p,3n)$ | 15.29 |
| $^{186}Re$ | 3.72 d | $^{186}W(p,n)$ | 1.37 |
| $^{183}Ta$ | 5.1 d | $^{186}W(p,\alpha)$ | 0.0 |
| | | $^{184}W(p,2p)$ | 7.74 |
| $^{184}Ta$ | 8.70 h | $^{186}W(p,n2p)$ | 15.11 |

Table 2 along with cross sectional data will be used to extract the proton energy spectrum from the samples exposed to the secondary beam at Livermore. There are several steps to relating the observed spectrum to the actual isotopes produced and finally to the energy of the incoming protons. The first involves a comparison of the various cross sections above threshold. Figure 1 shows a plot of the cross sections as a function of proton energy[6].

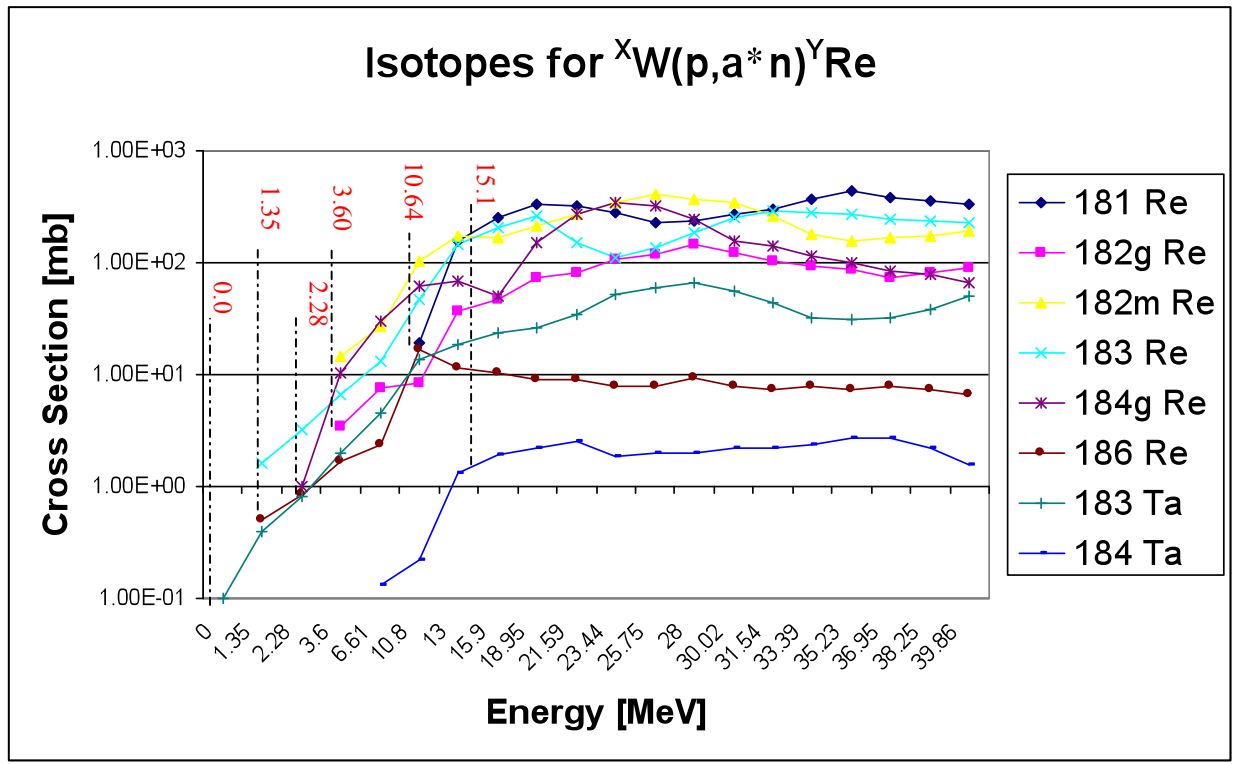

Figure 1: Cross sectional data for the production of radioactive isotopes on Tungsten.

These cross sections can be used to calculate a quantity called the yield, defined to be the total number of a given isotope produced per incoming proton as a function of proton energy. Equation (2) provides a numerical definition of yield.

$$yield = I_p \cdot \rho^* \cdot \frac{N_A}{m} \cdot t \cdot \sigma \qquad (2)$$

where $I_p$ is the number of incoming protons normalized to 1, $\rho^*$ is the density of the target (or 19.25 g/cm$^3$) weighted by the natural abundances in Table 1[(i)], $N_A$ is Avagadro's number, m is the gram molecular weight averaged as 184 g/mol, t is the target thickness and is 1mm for the collected data, and $\sigma$ is the cross section at a specified energy in MeV. Figure 2 shows the yields for each isotope listed above as a function of proton Energy.

(i) The abundance is used for only the threshold reaction, i.e. for $^{182}$Re the abundance of $^{182}$W is used.

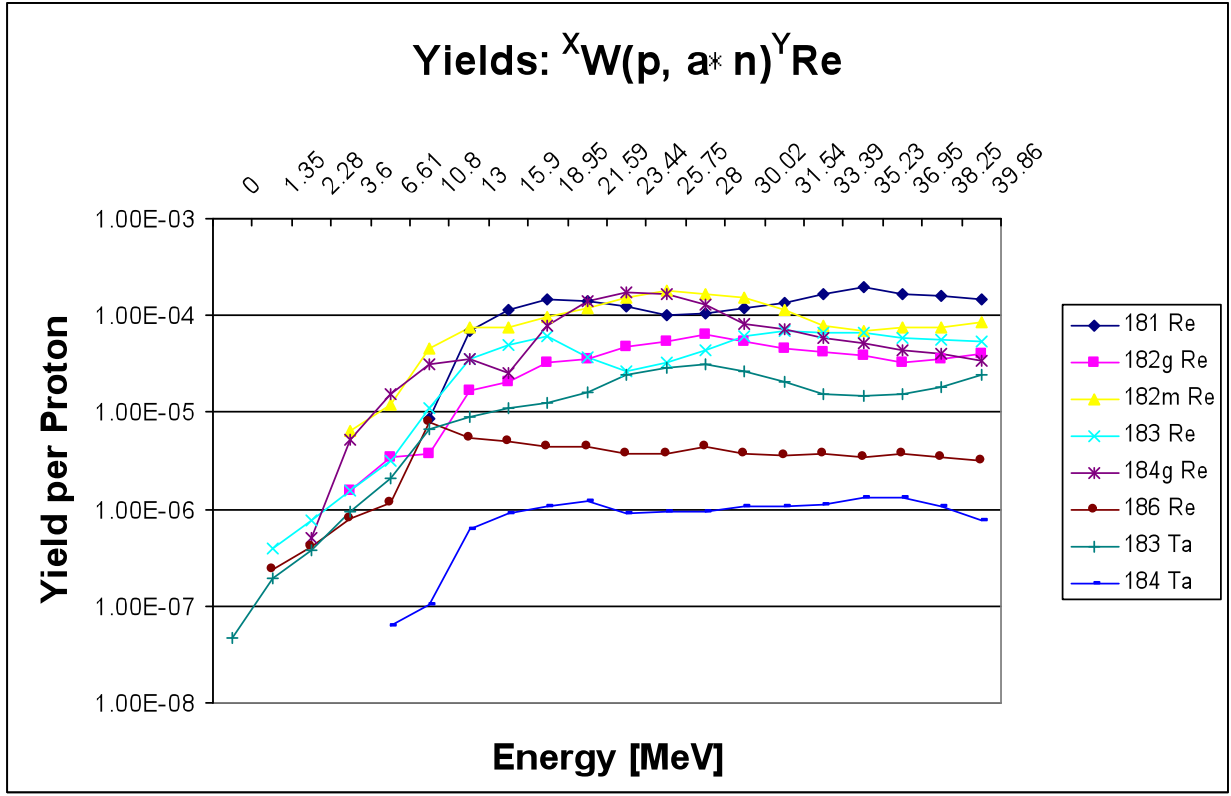

Figure 2: The yield as a function of the proton energy.

As with the cross section, the yield is a function of energy. To calculate the actual proton spectra, we will need to first fit then integrate the yield over some energy range. Intrator, Peterson and Zaidins used a five parameter function to fit yields of radioactive isotopes from zinc and cupper bombardment with protons[5]; as expressed in equation (3).

$$Yield(E) = [1 - \exp(F(E-Q)^{1/2})] x \exp(A + B\ln E + c\ln^2 E + D\ln^3 E] \qquad (3)$$

For simplicity, we use a polynomial function of the form given in equation (4).

$$Yield(E) = Ax^6 + Bx^5 + Cx^4 + Dx^3 + Ex^2 + Fx^1 + Gx^0 \qquad (4)$$

Table 3 lists the fit parameters. <u>NOTE</u>: going forward we follow the formalism of [5].

Table 3: Fit parameters for yields using a polynomial[ii]

| Isotope | A | B | C | D | E | F | G | $R^2$ |
|---|---|---|---|---|---|---|---|---|
| $^{183}$Ta | $2 \times 10^{-11}$ | $-1 \times 10^{-09}$ | $2 \times 10^{-08}$ | $-1 \times 10^{-07}$ | $5 \times 10^{-07}$ | $-3 \times 10^{-07}$ | $6 \times 10^{-08}$ | 0.9999 |
| $^{183}$Re | $-3 \times 10^{-10}$ | $1 \times 10^{-08}$ | $-1 \times 10^{-07}$ | $9 \times 10^{-07}$ | $-3 \times 10^{-06}$ | $5 \times 10^{-06}$ | $-2 \times 10^{-06}$ | 1.0000 |
| $^{186}$Re | $2 \times 10^{-10}$ | $-8 \times 10^{-09}$ | $1 \times 10^{-07}$ | $-1 \times 10^{-06}$ | $4 \times 10^{-06}$ | $-6 \times 10^{-06}$ | $4 \times 10^{-06}$ | 1.0000 |
| $^{184g}$Re | | $2 \times 10^{-10}$ | $-1 \times 10^{-08}$ | $2 \times 10^{-07}$ | $-2 \times 10^{-06}$ | $1 \times 10^{-05}$ | $-1 \times 10^{-05}$ | 1.0000 |
| $^{182g}$Re | | | $-2 \times 10^{-08}$ | $9 \times 10^{-07}$ | $-1 \times 10^{-05}$ | $6 \times 10^{-05}$ | $-1 \times 10^{-04}$ | 1.0000 |
| $^{182m}$Re | | | $-3 \times 10^{-08}$ | $1 \times 10^{-06}$ | $-1 \times 10^{-05}$ | $6 \times 10^{-05}$ | $-1 \times 10^{-04}$ | 1.0000 |
| $^{181}$Re | | | | $2 \times 10^{-07}$ | $-1 \times 10^{-05}$ | $2 \times 10^{-04}$ | $-1.2 \times 10^{-03}$ | 1.0000 |
| $^{184}$Ta | $5 \times 10^{-14}$ | $-1 \times 10^{-11}$ | $7 \times 10^{-10}$ | $-2 \times 10^{-08}$ | $4 \times 10^{-07}$ | $-3 \times 10^{-06}$ | $1 \times 10^{-05}$ | 0.9696 |

(ii) Fittings were performed over the ranges 0 – 15.11MeV and 15.11 – 40MeV. Parameters are for the lower range.

With the yield as a function of energy, we can calculate the numbers of radionuclides for a given reaction. Consider an incident proton beam with energy distribution ρ(E), the intensity of radioactivity can be written as in equation (5)

$$I_j = \int_0^\infty Y_j(E) \cdot \rho(E) dE \tag{5}$$

where Yj(E) is the yield function for the j$^{th}$ reaction. The next step is to divide the proton energy distribution into N regions, a histogram form, and rewrite it as in equation (6).

$$\rho(E) = \sum_{i=1}^{N} \frac{a_i \theta_i(E)}{E_i - E_{i-1}} \tag{6}$$

where $\theta_i$ is one in the energy region between $E_i$ and $E_{i-1}$ and zero everywhere else while $a_i$ is the particle flux in the i$^{th}$ energy bin.

The energy steps are values between the reaction thresholds. Table 4 gives the energy steps and the associated reaction with each higher step.

Table 4: Energy steps and the new reaction available

| Energy Step | Reactions Available |
|---|---|
| 0.0 – 1.35 | $^{183}$Ta |
| 1.35 – 1.37 | $^{183}$Re |
| 1.37 – 2.28 | $^{186}$Re |
| 2.28 – 3.60 | $^{184m}$Re, $^{184g}$Re |
| 3.60 – 10.64 | $^{182m}$Re, $^{182g}$Re |
| 10.64 – 15.11 | $^{181}$Re |
| 15.11 – 40.0 | $^{184}$Ta |

Now the j$^{th}$ reaction can also be rewritten in terms of this histogram function as in equation (7).

$$I_j = \sum_{i=1}^{N} a_i \int_{E_{i-1}}^{E_i} \frac{Y_j(E)}{E_i - E_{i-1}} dE = \sum_{i=1}^{N} a_i Y_{ij} = Y \cdot a \tag{7}$$

$Y_{ij}$ define a matrix with elements identifying each energy step and the associated reactions available. In the final form, Y·α, it can be seen that this matrix ($Y_{ij}$) converts a vector defining the proton flux ($\alpha_i$) to a vector defining the number of radionuclides ($I_j$) as a function of energy bin. We directly measure the numbers of radionuclides and thus directly extract $I_j$. Using the fits to the yields in Figure 2 (see Table 3) we can calculate the elements of $Y_{ij}$. Inverting this matrix, we can calculate $\alpha_i$, the proton flux as a function of energy. The next section covers the raw counts for the various radionuclide decays observed as well as a calculation of both the matrix and its inverse form.

## III. Tungsten Sample:

Before calculating the matrix and its inverse, we look at the spectrum taken for a tungsten plate exposed to a proton beam. The protons were produced following illumination of a foam target with the primary beam at the Jupiter facility. Figure 3 shows a sample of the gammas detected over a 12 hour period within 1 hr of exposure.

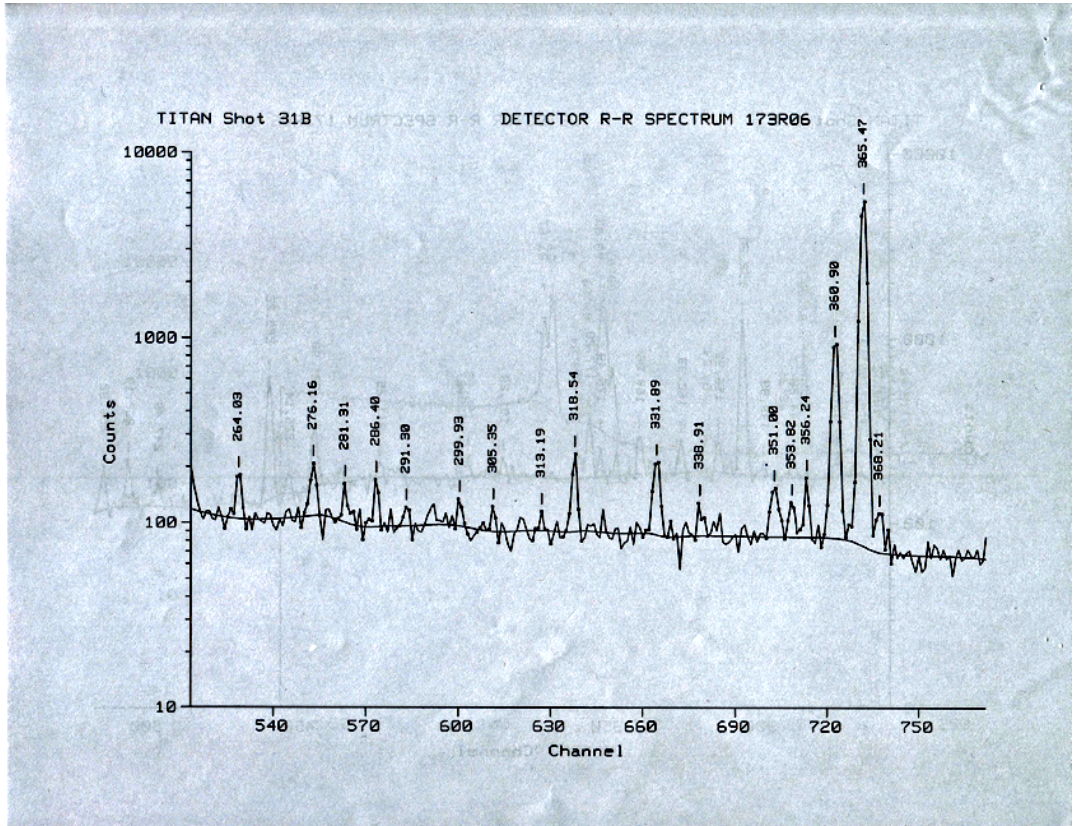

Figure 3: Sample spectrum for Tungsten showing gammas energies 250-370 keV.

The process of detecting radioactivity involves using a Germanium detector to accumulate gammas over some energy range. However, the raw counts, as shown here, can only be related to the numbers of radionuclides through a convolution involving the half-lift of the isotopes and a quantity called the branching fraction, the number of gammas emitted with a given energy per 100 decays. Equation (8) relates number of decays to the number of initial isotopes.

$$I_O = \varepsilon \cdot \Gamma \cdot I_j \exp(\frac{-t \cdot \ln 2}{\tau})$$

$$\int I_O dt = \varepsilon \cdot \int \Gamma \cdot I_j \exp(\frac{-t \cdot \ln 2}{\tau}) dt \qquad (8)$$

$$I_O t \Big|_{t=initial}^{t=final} = \varepsilon \cdot \Gamma \cdot (-\frac{\tau}{\ln 2}) \cdot I_j \exp(\frac{-t \cdot \ln 2}{\tau}) \Big|_{t=initial}^{t=final}$$

where $I_O$ represents the observed (raw) nuclide decays, $I_j$ represents the numbers of nuclides created, t give the amount of time over which the sample is observed or 12 hours, $\varepsilon$ represents the detector efficiency, $\Gamma$ gives the branching fraction associated with a given gamma (line) emitted, and $\tau$ is the half-life for the isotope. Inverting (8) gives equation (9).

$$I_j = \frac{I_O \cdot t \cdot \ln 2 \cdot \exp(\frac{-t \cdot \ln 2}{\tau})}{\varepsilon \cdot \Gamma \cdot (-\tau)} \quad (9)$$

This relationship can be used to extract the intensity with which each isotope was created and, ultimately, to determine the proton flux as a function of energy. Table 5 lists some of the gammas energies used to identify specific isotopes along with their branching fraction, half-lives, raw counts ($I_O$) and finally their respective intensity ($I_j$).

Table 5: Identifying gammas for each isotope[iii]

| Isotope | Emitted Gamma [keV] | Branching Fraction | Half-Life $\tau$ | Raw Counts | $I_j > E_{thres}$ – Per Gamma Line | Ave($I_j$) > $E_{thres}$ |
|---|---|---|---|---|---|---|
| $^{183}$Ta | 246.06 | 27.0 | 5.1 d | ~ 50 | 12 | 54.06 |
|  | 353.99 | 11.2 |  | 170 | 96 |  |
| $^{183}$Re | 162.32 | 23.3 | 70 d | 200 | 4 | 8.6 |
|  | 291.72 | 3.05 |  | 80 | 13 |  |
| $^{186}$Re | 137.16 | 9.42 | 3.72 d | 13 | 108 | 108.12 |
| $^{184m}$Re | 104.73 | 13.4 | 169 d | < 20 | 0.305 | 8.02 |
|  | 161.27 | 6.5 |  | < 20 | 0.629 |  |
|  | 216.55 | 9.4 |  | < 20 | 0.435 |  |
| $^{184g}$Re | 792.07 | 37.5 | 38.0 d |  |  |  |
|  | 894.76 | 15.6 |  | 269 | 16 |  |
|  | 903.28 | 37.9 |  |  |  |  |
| $^{182m}$Re | 1121.4 | 31.8 | 12.7 h |  |  | 2545 |
|  | 1189.2 | 15 |  | 2447 | 5550 |  |
|  | 1221.5 | 24.8 |  |  |  |  |
| $^{182g}$Re | 229.32 | 25.7 | 64.0 h | 1110 | 493 |  |
|  | 1121.3 | 22 |  |  |  |  |
|  | 1221.4 | 17.4 |  |  |  |  |
|  | 1231.0 | 14.9 |  | 2080 | 1593 |  |
| $^{181}$Re | 360.70 | 20.0 | 19.9 h | 2325 | 3199 | 4778 |
|  | 365.57 | 57.0 |  | 14,495 | 6997 |  |
|  | 639.30 | 6.4 |  | 962 | 4136 |  |
| $^{184}$Ta | 215.34 | 11.5 | 8.70 h | < 10 | 32 | 169 |
|  | 318.04 | 22.8 |  | 310 | 500 |  |
|  | 384.28 | 12.5 |  | ~ 40 | 118 |  |
|  | 414.01 | 72 |  | < 50 | 26 |  |
|  | 792.07 | 14.5 |  |  |  |  |
|  | 903.29 | 15 |  |  |  |  |

(iii) NOTE: Because lines can indicate several possible isotopes only lines unique to the isotope being considered are used to calculate $I_j$.

With $I_j$ established, all that is needed, to extract the proton flux, is the matrix elements calculated from the integrated yields. The elements of the yield matrix are simply an integration of the fit parameters over each energy step, from a lower threshold value to the next higher one (see Table 4), for each reaction. Table 6 gives the elements of the $Y_{ij}$.

Table 6: Yij Matrix elements

| Energy Steps ΔE [MeV] | Integrated Yield Fits | | | | | | | I val |
|---|---|---|---|---|---|---|---|---|
| | $^{183}$Ta | $^{183}$Re | $^{186}$Re | $^{184m}$Re $^{184g}$Re | $^{182m}$Re $^{182g}$Re | $^{181}$Re | $^{184}$Ta | |
| 0.0 – 1.35 | 1.516E-07 | | | | | | | 1 |
| 1.35 – 1.37 | 7.783E-09 | 2.764E-06 | | | | | | 2 |
| 1.37 – 2.28 | 7.214E-07 | 1.564E-06 | 1.141E-06 | | | | | 3 |
| 2.28 – 3.60 | 3.046E-06 | 5.882E-06 | 1.526E-06 | 3.612E-05 | | | | 4 |
| 3.60 – 10.64 | 2.078E-04 | 9.626E-04 | 3.401E-04 | 1.327E-03 | 7.913E-04 | | | 5 |
| 10.64 – 15.11 | 7.075E-04 | 4.220E-03 | 4.065E-04 | 1.346E-04 | 1.441E-03 | 1.459E-03 | | 6 |
| 15.11 – 40.0 | 2.590E-01 | 17.02E-00 | 2.487E-02 | 5.375E-00 | 1.124E-01 | 1.080E-01 | 1.069E-03 | 7 |
| j val | 1 | 2 | 3 | 4 | 5 | 6 | 7 | |

Note: The first and last columns involve isotopes of Tantalum which could have come from gamma induced reactions on $^{184}$W and/or $^{186}$W. In a second note the gamma spectrum is extracted using the produced isotopes of Ta. To justify not including these effects here, consider the gamma induced transitions that will create $^{183}$Ta and $^{184}$Ta shown in equation (10) below.

$$^{184}W + \gamma \rightarrow \,^{183}Ta + p$$
$$^{186}W + \gamma \rightarrow \,^{183}Ta + p2n \qquad (10)$$
$$^{186}W + \gamma \rightarrow \,^{184}Ta + pn$$

These transitions require energies well above the threshold to produce the same Ta isotopes using protons on naturally occurring W. Therefore, these contributions to the cross section will be evaluated separately and are omitted when considering the proton effects.

Finally, the matrix $Y_{ij}$ must be inverted ($Y_{ij}^{-1}$) to obtain the proton flux from the intensities of each reaction $I_j$. This process of inverting a matrix can readily be found in a number of textbooks on linear algebra[7]. Equation (11) shows the basic idea behind matrix inversion.

$$[A_{ij}][A_{ij}^{-1}] = \begin{bmatrix} 1 & 0 \\ 0 & 1 \end{bmatrix} = [I] \qquad (11)$$

Here [ I ] is the identity matrix. The matrix in equation (9) is in a form commonly referred to as a lower triangular form; only elements along or below the diagonal are nonzero. This simplifies the process of finding an inverse considerably. Consider the 3x3 matrix and its inverse in equation (12); a clear pattern can be seen.

$$\begin{bmatrix} a & 0 & 0 \\ b & d & 0 \\ c & e & f \end{bmatrix} \begin{bmatrix} \dfrac{1}{a} & 0 & 0 \\ \dfrac{-b}{ad} & \dfrac{1}{d} & 0 \\ \dfrac{be-cd}{adf} & \dfrac{-e}{df} & \dfrac{1}{f} \end{bmatrix} = \begin{bmatrix} 1 & 0 & 0 \\ 0 & 1 & 0 \\ 0 & 0 & 1 \end{bmatrix} \qquad (11)$$

For the inverse matrix, the diagonal elements are one divided by each diagonal element of the original matrix. Like the original matrix, the inverse is also a lower triangular matrix with off-axis elements determine by combinations of those of the original matrix.

Table 7 gives the elements for the inverse Yij matrix.

Table 7: $Y_{ij}^{-1}$ Matrix elements

| | | | | | | | |
|---|---|---|---|---|---|---|---|
| 6.60E+06 | | | | | | | 1 |
| -1.86E+04 | 3.62E+05 | | | | | | 2 |
| -4.14E+06 | -4.96E+05 | 8.76E+05 | | | | | 3 |
| -3.78E+05 | -3.79E+04 | -3.70E+04 | 2.77E+04 | | | | 4 |
| 7.06E+05 | -1.63E+05 | -3.14E+05 | -4.64E+04 | 1.26E+03 | | | 5 |
| 3.35E+07 | -7.44E+05 | 7.00E+04 | 4.33E+04 | -1.25E+03 | 6.85E+02 | | 6 |
| -2.77E+09 | -5.46E+09 | 1.92E+08 | -1.39E+08 | -6.73E+03 | -6.92E+04 | 9.35E+02 | 7 |
| 1 | 2 | 3 | 4 | 5 | 6 | 7 | |

Now with the inverse can be applied to equation (7) and, using the nuclide intensity vector extracted in Table 5, the proton flux as a function of energy can be found.

**IV Conclusion**

The process of determining the spectrum of protons produced from a foam target illuminated with the Titan laser, using existing methodology, has proven a straightforward one. The extracted spectrum can now be used to estimate backgrounds to further experimentation at the Jupiter facility. Furthermore, this technique will be applied to data from irradiated samples to extract the gamma and electron spectra. Thus the backgrounds to future measurements at Jupiter will be completely characterized.

**V. Acknowledgements**

This work was made possible through the generosity of staff at the LLNL's Jupiter Laser Facility. Robert Cauble made possible the collection of the initial samples following exposure and Hui Chen allowed myself and Jerry Peterson to participate in a scheduled run. A special thanks must also be given to Dr. Elliot Treadwell, of Florida A&M University, who allowed the second set of spectrum to be taken on his laboratory's Germanium detector.

**References:**

[1] Chen et. al., Relativistic Positron Creation Using Ultra-Intense Short Pulse Lasers, PRL 102 105001 (2009)

[2] F. Reines and C. Cowan, Free Antineutrino Absorption Cross Section., Phys. Rev. 113, 273-279 (1959)

[3] The MiniBooNE Collaboration, Antineutrino Oscillation Results, Phys Rev. Lett. 103, 111801 (2009)

[4] J. Cork et. al., The Radioactive Decay of Tungsten 181, Phy. Rev. 92, 1 (1953)

[5] T. Intrator, R Peterson, and C. Zaidins, Determination of Proton Spectra By Thick Target Radioactive Yields, Nuclear Instruments and Methods 188 (1981) 347-352

[6] M.U. Khandaker, et al…., Excitation functions of proton induced nuclear reactions on $^{nat}$W up to 40 MeV

[7] G. Strang, Linear Algebra and Its Applications